\documentclass[jair,twoside,11pt,theapa]{article}
\usepackage[utf8]{inputenc}
\usepackage[T1]{fontenc}
\usepackage{graphicx}
\usepackage{subfigure}
\usepackage{float}
\usepackage[section]{placeins}

\usepackage{algorithm}
\usepackage{algorithmic}

\usepackage{lipsum}
\usepackage{hyperref}
\usepackage{amsmath}
\usepackage{amssymb}
\usepackage{color}
\usepackage{jair, theapa, rawfonts}

\ShortHeadings{Exploiting a comparability mapping to improve bi-lingual data categorization}
{Marteau, \& Ke}
\firstpageno{1}

\begin{document}

\title{Exploiting a comparability mapping to improves bi-lingual data categorization:  a three-mode data analysis perspective}

\author{\name Pierre-Francois Marteau \email pierre-francois.marteau@univ-ubs.fr \\
		\addr IRISA (UMR CNRS 6074), Universite de Bretagne Sud, 56000 Vannes, France
		\AND
		\name Guiyao Ke \email yannick.crystal@gmail.com \\
        \addr IRISA (UMR CNRS 6074), Universite de Bretagne Sud, 56000 Vannes, France}

\maketitle

%\markboth{Draft paper in submission, October~2013}%
%{Shell \MakeLowercase{\textit{et al.}}: Bare Demo of IEEEtran.cls for Computer Society Journals}

%\IEEEcompsoctitleabstractindextext{%
\begin{abstract}
 We address in this paper the co-clustering and co-classification of bilingual data laying in two linguistic similarity spaces when a comparability measure defining a mapping between these two spaces is available. A new approach that we can characterized as a three-mode data analysis scheme,  is proposed to mix the comparability measure with the two similarity measures.  Our  aim is to improve jointly the accuracy of classification and clustering tasks performed in each of the two linguistic spaces, as well as the quality of the final alignment of comparable clusters that can be obtained. We used first some purely synthetic random data sets to assess our formal similarity-comparability mixing model. We then propose two variants of the comparability measure that has been defined by \cite{LiGaussier2010} in the context of bilingual lexicon extraction to adapt it to clustering or categorizing tasks. These two variant measures are subsequently used to evaluate our similarity-comparability mixing model in the context of the co-classification and co-clustering of comparable textual data sets collected from Wikipedia categories for the English and French languages. Our experiments show clear improvements in clustering and classification accuracies when mixing comparability with similarity measures, with, as expected, a higher robustness obtained when the two comparability variant measures that we propose are used. We believe that this approach is particularly well suited for the construction of thematic comparable corpora of controllable quality.      

\end{abstract}

%\begin{keywords}
%Comparable corpora, Comparability measures, Classification, Clustering, Cluster mapping
%\end{keywords}}

%\IEEEdisplaynotcompsoctitleabstractindextext

%\IEEEpeerreviewmaketitle

%\title{Improving the clustering or categorization of bi-lingual data by means of comparability mapping}

%\motsClefs{Comparable corpora, Comparability measures, Classification, Clustering, Clusters mapping}

\section{Introduction}

Parallel corpora are sets of tuples of aligned documents that are formed with texts placed alongside with their translation(s). If such resources are of great utility in particular in the field of assisted translation or multilingual information retrieval, they are expensive to develop and often difficult to transpose from a specialty domain to another. The notion of comparable corpora has emerged in the nineties to palliate this lack of versatility and expensiveness and to offer avenues to a wider scope of applications such as multilingual terminology extraction, multilingual information retrieval or knowledge engineering \cite{Baker1996}, \cite{EAGLES1996}. However, the notion of comparability between documents expressed in different languages is not easy to introduce: it is widely admitted that two documents in different languages are comparable when they share analogous criteria of composition, genre and topics. The term of comparable corpora was introduced by \cite{Fung1998}, \cite{MunteanuFM04} and remains quite subjective. \cite{Dejean2002} proposed a quantitative definition of the concept of comparability according to which "Two corpora in two languages $\mathcal{L}_1$ and $\mathcal{L}_2$ are called comparable if there is a significant sub-part of the vocabulary of the $\mathcal{L}_1$ language corpus, respectively $\mathcal{L}_2$ language corpus, whose translation is in the corpus of language $\mathcal{L}_2$, respectively $\mathcal{L}_1$." \cite{LiGaussier2010} have then derived a quantitative measure that is based on a bilingual translation dictionary. This measure consists primarily in counting the presence of the translations of dictionary entries that occur in the paired documents. It depends in a non-explicit way upon jointly the coverage of the bilingual translation dictionary and the studied corpora themselves.

This comparability measure defined for bilingual corpora indeed applies when dealing with monolingual documents that partition in two distinct linguistic spaces, as far as a bilingual dictionary connecting the two spaces is available. At a document level we thus face a situation where monolingual similarity measures exist in each linguistic space that are potentially linked by a comparability measure.  In the scope of the construction of thematic comparable corpora, this leads to address the co-classification or co-clustering of bilingual data since we are targeting the mapping of highly \textit{comparable} clusters of documents that are furthermore thematically coherent in each linguistic space, i.e. characterized by a high \textit{intra-similarity}. 
We confront such situation when harvesting multilingual data from the web for instance. With the need for comparable resources getting pressing, approaches that exploit consistently similarity and comparability measures are becoming particularly useful.

There is apparently no existing direct method available to map comparable clusters of documents that lay in two different linguistic spaces. Nevertheless, there exist some work which is somehow related to this problem, like biclustering, co-clustering, or two-mode clustering introduced by \cite{MirkinBoris96} and \cite{VanMechelenBockDeBoeck04}. However, these works are mainly relevant to the clustering of the rows and columns (instances and features axes) of a given matrix and does not fit with the sort of three-mode categorization or clustering we are facing. 

Recently, \cite{Jagarlamudi:2011ACL}, \cite{Jagarlamudi:2011EMNLP} have developed quite successfully a supervised method that learns interlingual representations from aligned training documents. They exploit word association measures and bilingual dictionary to remove noisy pairs of aligned documents. In \cite{AminiGoutte2010} the authors proposed to learn a co-classification from multi-lingual corpora, based on a co-regularization of the categories in order to maintain a consistency of the categorization process across languages. \cite{LiGaussier2011} have proposed a solution for clustering bilingual corpora by using the comparability measure only.  

However, if our approach also seeks the joint clustering or classification of data that lay in two distinct linguistic spaces, it aims at exploiting, in conjunction with a comparability mapping existing between the two spaces, \textit{native} similarity measures (a native similarity has to be understood as any quantitative intra-language similarity measure, such as a cosine similarity measure) existing within these two linguistic spaces. More precisely, the proposed approach is lying between the work reported in \cite{AminiGoutte2010} and \cite{LiGaussier2011}. It exploits directly, i.e. without any learning phase,  the comparability measure that maps the two linguistic spaces to provide new similarity measures that combine \textit{native} similarity measures with a similarity measure that is \textit{induced} by the comparability mapping. 

Thus the approach that we develop in the following sections only rely on a bilingual dictionary and does not assume that any aligned data preexist as learning data. Indeed, this approach could be enriched using a feature-extraction technique, such as the one proposed in \cite{Vu:2009} for instance, to align bilingual documents that have a similar content.

After introducing our main motivations, we develop a straightforward mixing model to combine similarity and comparability measures in an efficient way that allows for the development of consistent co-clustering and co-classification of comparable data and assess it on purely synthetic data. We then address the concept of comparability for mapping bilingual textual data, and define, from the original measure proposed by \cite{LiGaussier2011}, two alternative variant measures to overcome some limitation of the original measure. To assess the proposed approach on real textual data, we then detail an experimentation based on a subset of comparable documents collected from some Wikipedia categories. Basically, we evaluate jointly the three tested comparability measures and the proposed similarity-comparability mixing model in the scope of co-classification and co-clustering of bilingual data. Finally we discuss our results and draw some perspectives.

\section{Motivations: similarity spaces connected by a comparability mapping}
\begin{figure}[!]
  \centering
  \includegraphics[scale=0.4, angle=0]{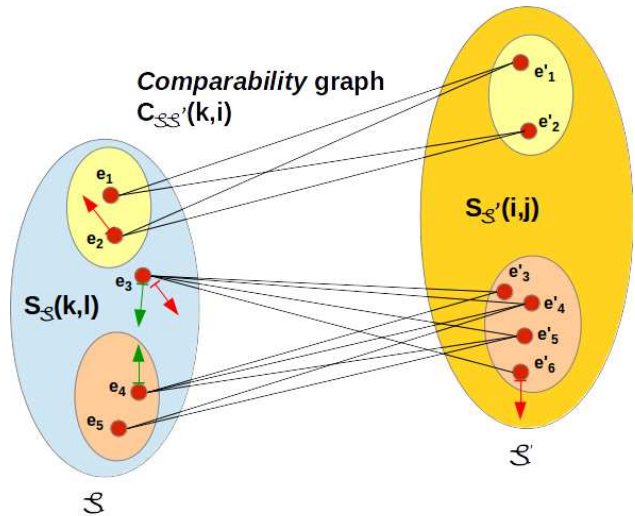}
  \caption{Two similarity spaces connected by a comparability mapping.}
  \label{fig:motivations}
\end{figure}

When confronting with complex data one may encounter situations where two distinct spaces $\mathcal{S}$ and $\mathcal{S}'$, in which preexist \textit{native} similarity measures $S_{\mathcal{S}}$ and $S_{\mathcal{S}'}$, are interconnected by a mapping $C_{\mathcal{S}\mathcal{S}'}$. Figure \ref{fig:motivations} gives an example of such situation. This is the case when considering comparable corpora that are composed with texts written in at least two distinct languages. For such data, a bilingual dictionary allows for the construction of a comparability measure \cite{LiGaussier2010} yielding to the definition of a comparability mapping \cite{MarteauMenier2013} that links the two sets of comparable documents. More generally speaking, such case arises in situations where heterogeneous but analogous data is available, through different sources, in different formats, or characterized using different sets of descriptors,  or comply to different semantic models such as heterogeneous ontologies for instance, etc.  The principle of mapping heterogeneous but comparable data that we address is quite general since it takes the form of any bipartite weighted undirected graph. We call it a comparability mapping. Hence, a comparability mapping establishes a bi-directional connection between the elements of the two similarity spaces that could be used to challenge native similarity measures (or distances) defined in the two spaces. By doing so, we introduce a kind of three mode data analysis scheme: the two first modes are associated to the two \textit{native} similarity spaces, while the third mode is related to the comparability mapping itself that links these two spaces. 

As an example, in figure \ref{fig:motivations}, two discrete sets of elements $\mathcal{S}$ and $\mathcal{S}'$ are presented. We suppose that the notion of \textit{native} similarity between elements of these sets is defined, we call them respectively $S_{\mathcal{S}}(.,.)$: $\mathcal{S} \times \mathcal{S} \rightarrow \mathbb{R}$  and $S_{\mathcal{S}'}(.,.)$: $\mathcal{S}' \times \mathcal{S}' \rightarrow \mathbb{R}$. \\
Furthermore, the two sets are point-wisely connected by a mapping defined by a comparability measure $C_{\mathcal{S}\mathcal{S}'}(.,.)$: $\mathcal{S} \times \mathcal{S}' \rightarrow \mathbb{R}$. This mapping  that takes the form of a bipartite graph is a comparability mapping. The edges of this graph are bidirectional and weighted  by a real value that can be bounded into $[-1,1]$.  

The main idea that we develop in this article is that of similarity \textit{induced} by a comparability mapping: in other words, if two elements in the set $\mathcal{S}$  are mapped to a same subset of elements in the set $\mathcal{S}'$, then their similarity should be important (and vice versa). \textit{a contrario}, if two elements in the set $\mathcal{S}'$  are mapped to disjoint subsets of elements in the set $\mathcal{S}$, then their similarity should be small (and vice versa). Thus, in figure \ref{fig:motivations}, from the point of view of the similarity derived from the comparability mapping alone, element $e_3$ should move away from element $e_2$ to get closer to elements $e_4$ and $e_5$. Similarly, element $e'_6$ should move away from elements $e'3$, $e'_4$ and $e'_5$. The expected utility of such a similarity \textit{induced} by a comparability mapping is a kind of noise filtering capability. When exploited in conjunction with \textit{native} similarity measures in $\mathcal{S}$  and $\mathcal{S}'$ a fusion of complementary sources of knowledge is achieved that could help building more robust similarity functions into $\mathcal{S}$ and $\mathcal{S}'$ spaces.  The noise in question could have many sources, in particular it could be inherent to the representational models of the element themselves due to a lack of knowledge, e.g. lack of structural variability, data heterogeneity, semantic ambiguities, etc.

\section{Combining similarity and comparability: a three-mode analysis scheme}

\subsection{Similarity measure \textit{induced} by a comparability mapping}
In this line of work, \cite{MarteauMenier2013} proposed an algorithm, \textit{Hit-ComSim}, to iteratively construct the concept of similarity \textit{induced} by a comparability bipartite graph. Unfortunately, this algorithm does not scale well due to its high algorithmic complexity (in $O(N^{4})$). We propose here a much more straightforward approach that consists in  exploiting directly the comparability matrix constructed from the two bilingual finite collections of documents.

Let us consider $\mathcal{S}$ and $\mathcal{S}'$ two collections of documents belonging to two distinct linguistic spaces ($\mathcal{L}$ and $\mathcal{L}'$ respectively) in which two \textit{native} similarity measures $S_{\mathcal{S}}$ and $S_{\mathcal{S}'}$ are defined. Let $C(.,.): S_{\mathcal{S}} \times S_{\mathcal{S}'} \rightarrow \mathcal{R}$  be the comparability function that maps the two finite collections or equivalently that defines a weighted bipartite graph between the two linguistic spaces. The two similarity functions $S_{\mathcal{S}}$, $S_{\mathcal{S}'}$ and the comparability measure $C$ allows for the definition of the following three-mode analysis scheme.

We define the similarity measures \textit{induced} by the comparability mapping $C$ as the following two normalized (in $[-1,1]$) measures respectively noted $S_{\mathcal{S}_{1}, C}$  and $S_{\mathcal{S}', C}$:\\

$\forall (d_i, d_j) \in  \mathcal{S}^{2}$ and $\forall (d'_i, d'_j) \in  \mathcal{S}'^{2}$ 
\begin{eqnarray}
\begin{array}{ll}
S_{\mathcal{S}, C}(d_i, d_j)=\dfrac{CC^{T}(i,j)}{\sqrt{CC^{T}(i,i)CC^{T}(j,j)}}\\
S_{\mathcal{S}', C}(d'_i, d'_j)=\dfrac{C^{T}C(i,j)}{\sqrt{C^{T}C(i,i)C^{T}C(j,j)}}
\end{array}
\end{eqnarray}

The interpretation of the similarity measures that are \textit{induced} by a comparability mapping $C$ is straightforward. First, considering each row $i$ of the $C$ matrix as a feature vector that characterizes document $d_i \in \mathcal{S}$, for any $(d_i,d_j) \in \mathcal{S}$, $CC^{T}(i,j)$ can be interpreted as an inner product between the two feature vectors representing $d_i$ and $d_j$ respectively. Then,  $S_{\mathcal{S}, C}(d_i, d_j)$ is nothing but a cosine similarity between documents $d_i$ and $d_j$ based on the comparability mapping only. 

Similarly, considering each column $i$ of the $C$ matrix as a feature vector that characterizes document $d'_i \in \mathcal{S}'$, $S_{\mathcal{S}', C}(d'_i, d'_j)$ is nothing but a cosine similarity between documents $d'_i$ and $d'_j$ $\in \mathcal{S}'$ based on the comparability mapping only.

\subsection{Mixing \textit{native} similarity and \textit{induced} similarity}
The comparability/similarity mixing model that we propose is a simple linear combination of the \textit{native} and \textit{induced} similarity measures defined in each linguistic space.   Basically we use a single parameter $\alpha \in [0,1]$ to combine linearly the two measures as follows
\begin{eqnarray}
\begin{array}{ll}
S'_{\mathcal{S}}(d_i, d_j)=\alpha S_{\mathcal{S}, C}(d_i, d_j) + (1 - \alpha)S_{\mathcal{S}}(d_i, d_j) \\
S'_{\mathcal{S}'}(d'_i, d'_j)=\alpha S_{\mathcal{S}', C}(d'_i, d'_j) + (1 - \alpha)S_{\mathcal{S}'}(d'_i, d'_j) 
\end{array}
\label{eq:mixedSimilarities}
\end{eqnarray}

Since the \textit{induced} similarity measures are normalized into the interval $[-1,1]$, we advocate using a cosine similarity as \textit{native} similarity measures in the two connected  linguistic spaces such that the mixed similarity measures defined by equation \ref{eq:mixedSimilarities} are consistent.  

Finally, as this model mixes two sources of native similarity with the induced similarity measures that are directly derived from the comparability mapping, it implements the so-called three-mode data analysis scheme that we were referring to in the motivation section. 

\section{Experimenting on synthetic data}
To evaluate the effectiveness of the proposed similarity-comparability mixing model, we generated 20 distinct tests by randomly defining:
\begin{itemize}
\item two similarity spaces $\mathcal{S}$ and $\mathcal{S}'$,  
\item a categorization of the elements within each of these spaces, 
\item the comparability mapping between them.
\end{itemize} 

 The algorithm \ref{Algo:rand} describes the way each of these 20 tests is generated. The variance parameters $V_s$ and $V_c$ are set up such that the categories are significantly overlapping making the classification problems difficult enough. To put more discriminative weight on the \textit{native} similarity measures, the variance $V_c$ associated to the comparability mapping matrix that is used to provide the \textit{induced} similarity measures is three times the variance $V_s$ used for producing the \textit{native} similarity measures.

\begin{algorithm}
\caption{Random generation of two \textit{native} similarity spaces connected by a comparability mapping.  The algorithm provides two random similarity matrices $S_{\mathcal{S}}$ and $S_{\mathcal{S'}}$, the random comparability mapping matrix $C_{\mathcal{S},\mathcal{S}'}$,  two sets of comparable clusters associated to a cluster map, $mapC$, also randomly defined on spaces $\mathcal{S}$ and $\mathcal{S}'$.}
\label{Algo:rand}
\begin{algorithmic}
\STATE // $Randn(n,m)$ returns an n-by-m matrix containing pseudo-random values drawn
\STATE // \hspace{2mm} from a normal distribution with mean zero and standard deviation one.
\STATE // $Randn()$ returns a single value from the previous distribution.
%\REQUIRE \texttt{$n>0$, $\pi \in \mathcal{M}_n$, $A \in \mathbb{U}_n$}
\STATE 0) $V_s=1.0;$ $V_c=3.0;$
\STATE 1) Randomly select the number of clusters $n_\mathcal{S}^c$ (resp. $n_{\mathcal{S}'}^c$) in $\mathcal{S}$ (resp. $\mathcal{S}'$) from the set $\{3,..., 18\}$;
\STATE 2) For each cluster $c_k$ in $\mathcal{S}$ (resp. $c'_l$ in $\mathcal{S}'$) randomly select the number of elements in $c_k$, $|c_k|$ (resp.$c'_l$, $|c'_l|$) from the set $\{20,.., 40\}$;
\STATE 3) For each pair of elements $(e_i, e_j)$ in $\mathcal{S}^2$ (resp. $\mathcal{S}'^2$)
	\IF{$e_i$ and $e_j$ belong to the same cluster}
		\STATE $S_{\mathcal{S}}(e_i,e_j)=0.5 + V_S*Randn()$;
		\STATE resp. $S_{\mathcal{S}'}(e_i,e_j)=0.5 + V_S*Randn()$;
	\ELSE
		\STATE $S_{\mathcal{S}}(e_i,e_j)=-0.5 + V_S*Rand()$;
		\STATE resp. $S_{\mathcal{S}'}(e_i,e_j)=-0.5 + V_S*Rand()$;
	\ENDIF
\STATE 4) $mapC = Randn(n_\mathcal{S}^c, n_{\mathcal{S}'}^c)$;
\STATE $J=0$
        \FOR{$k=1:n_\mathcal{S}^c$}
        	\STATE $I=0$;
            \FOR{$l=1:n_{\mathcal{S}'}^c$}
                \FOR{$i=1:|c'_l|$}
                    \FOR{$j=1:|c_k|$} 
                        \STATE $C_{\mathcal{S},\mathcal{S}'}(I+i,J+j)=randn()*Vc+mapC(l,k)$;
                    \ENDFOR
                \ENDFOR
                \STATE $I=I+|c'_l|$;
            \ENDFOR
            \STATE $J=J+|c_k|$;
         \ENDFOR    
\end{algorithmic}
\end{algorithm}

Table \ref{tab:synClust} gives for each similarity spaces $\mathcal{S}$ and $\mathcal{S}'$ the number of elements and the number of clusters for each of the 20 tests.
\begin{table}[!]
\center
\setlength{\tabcolsep}{-0pt}
\begin{tabular}{|c|c|c|c||c|c|c|c|}
  \hline
   \multicolumn{2}{|c|}{$\mathcal{S}$} &  \multicolumn{2}{|c|}{$\mathcal{S}'$} & \multicolumn{2}{|c|}{$\mathcal{S}$} &  \multicolumn{2}{|c|}{$\mathcal{S}'$}\\
   \hline
    \#clusters & \#elements  & \#clusters & \#elements & \#clusters & \#elements  & \#clusters & \#elements\\
  \hline
  \hline
15	& 491	& 7	& 189 & 15	& 411	& 11	& 314 \\
12	& 355	& 5	& 148 & 15	& 524	& 12	& 343 \\
16	& 485	& 10	& 309 & 6	& 169	& 17	& 488\\
16	& 516	& 8	& 243 & 8	& 223	& 14	& 420 \\
9	& 287	& 17	& 486 & 18 	& 546	& 5	& 147\\ 
8	& 247	& 11	& 322 & 13	& 415	& 12	& 385 \\
18	& 540	& 14	& 405 & 11	& 347	& 16	& 499 \\
14	& 445	& 12	& 386 & 11	& 366	& 6	& 191\\
8	& 246	& 17	& 540 & 17	& 547	& 17	& 528 \\
18	& 565	& 4	& 136 & 5	& 160	& 16	& 456 \\
 \hline
\end{tabular}
\caption{Number of clusters and number of total elements in each synthetic similarity spaces $\mathcal{S}$ and $\mathcal{S}'$ for the twenty tests used for this experiment.}
\label{tab:synClust}
\end{table}

\begin{figure}[!]
  \centering
  \begin{tabular}{ccc}
	\includegraphics[scale=0.7]{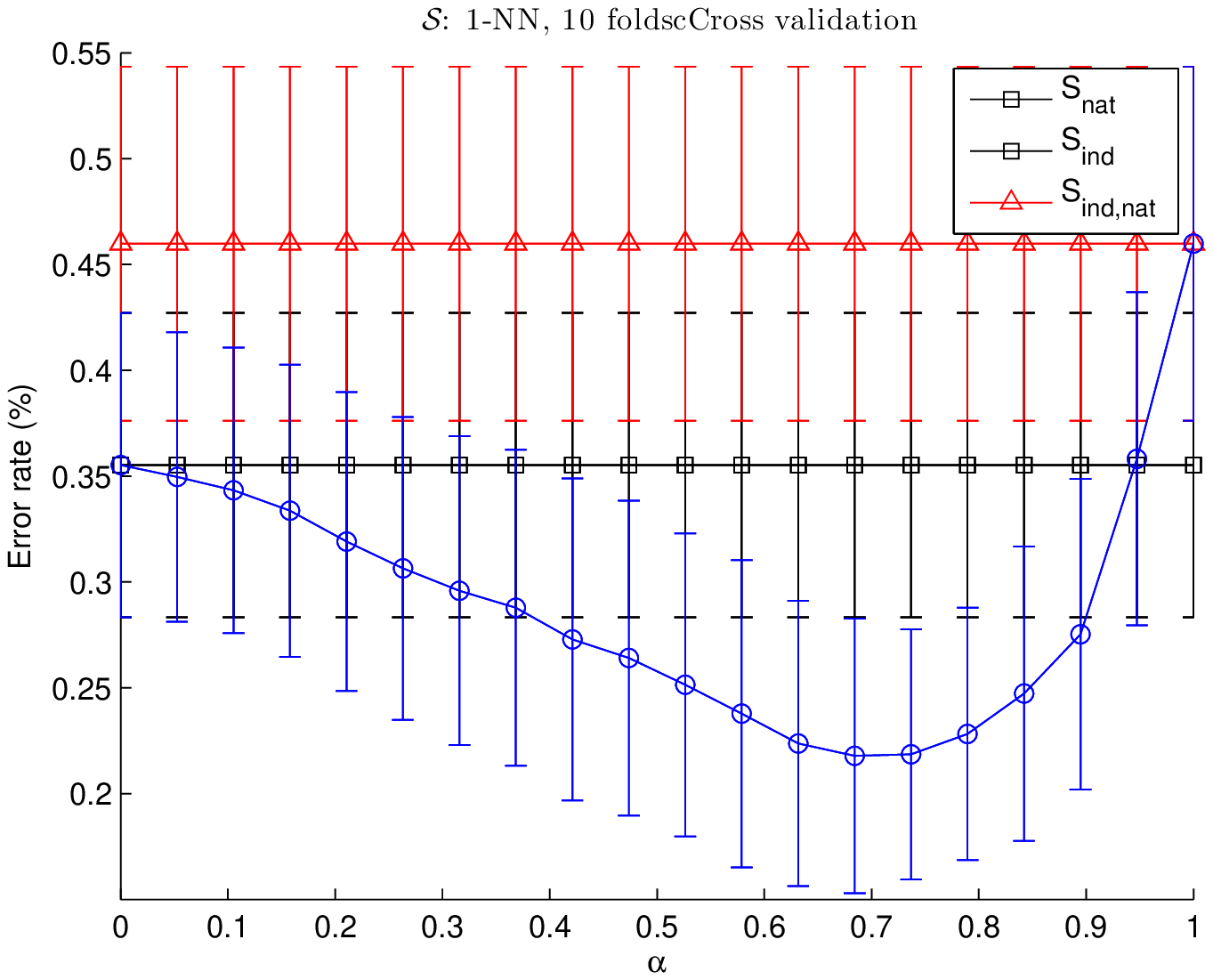} \\
	\includegraphics[scale=0.7]{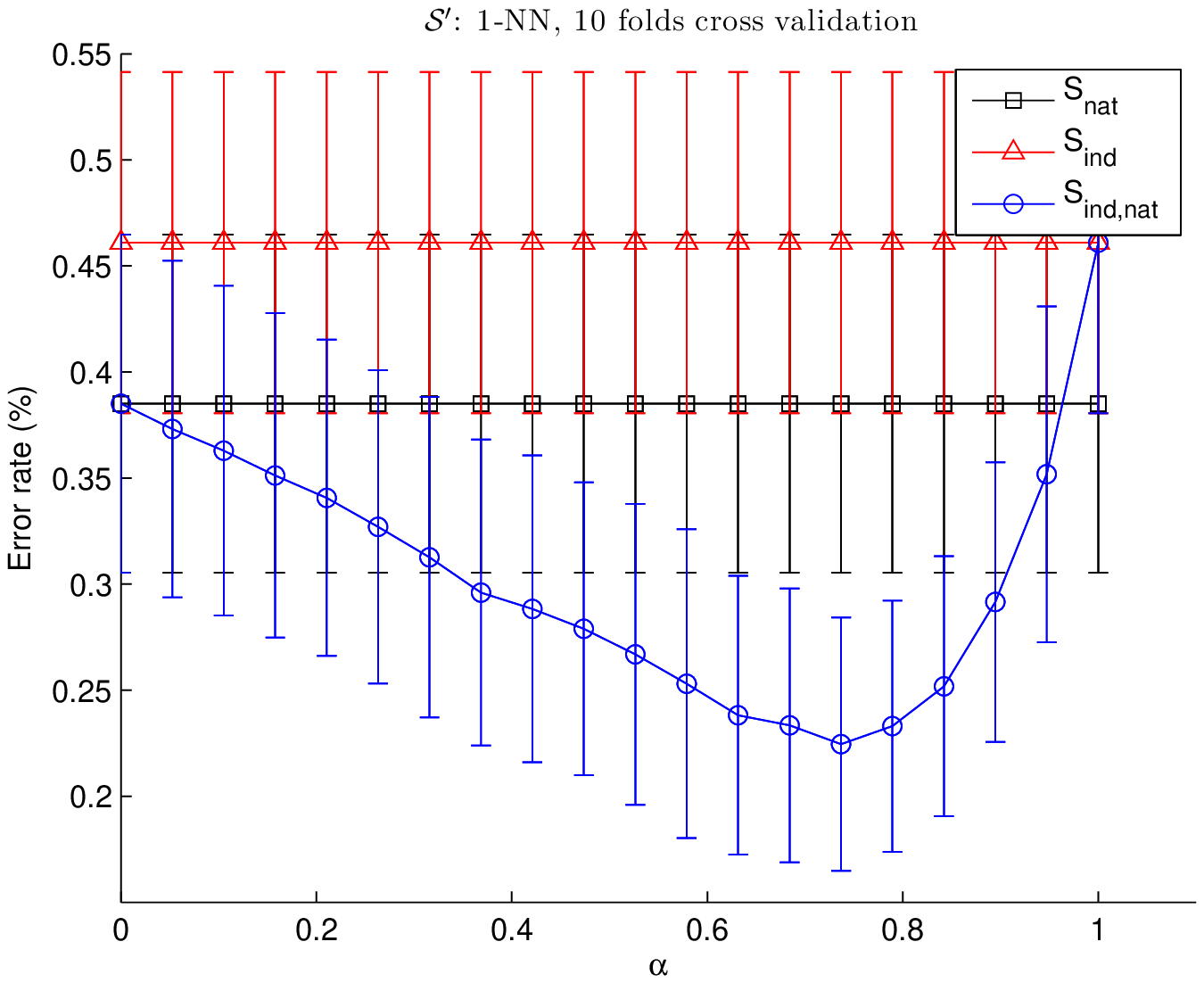} \\
  \end{tabular}
  \caption{Comparability/similarity mixing effect on the 1-NN classification task, according to a 10-fold cross validation procedure. The error rates in \% %(top) and F1 measure (bottom) 
  are given when the \textit{native} similarity alone is used (black square curve) when the \textit{induced} similarity alone is used (red triangle curve) and when the  mixing model is used as a function of parameter $\alpha \in [0,1]$ (blue circle curve).  Top $\mathcal{S}$, bottom $\mathcal{S}'$ similarity spaces}
	\label{fig:effetmixknn1}
\end{figure}

%\begin{figure}[!h]
%  \centering
%  \begin{tabular}{ccc}
%    % Requires \usepackage{graphicx}
%	\includegraphics[scale=0.5]{o1-1NN-F1.eps}&
%	\includegraphics[scale=0.5]{o2-1NN-F1.eps} \\
%  \end{tabular}
%  \caption{Comparability/similarity mixing effect on the 1-NN classification task, according to a 10-fold cross validation procedure. F1 measure \% is given when the native similarity alone is used (black square curve) when the induced similarity alone is used (red triangle curve) and when the  mixing model is used as a function of parameter $\alpha \in [0,1]$ (blue circle curve).  Left $\mathcal{S}$, right $\mathcal{S}'$ similarity spaces}
%	\label{fig:effetcompknn1}
%\end{figure}

To evaluate the effectiveness of a 1-NN classification as the mixing parameter $\alpha$ varies, we use the classification error rate measure. 
%and the F1 score defined as $F_1 = 2 \cdot \frac{\mathrm{precision} \cdot \mathrm{recall}}{\mathrm{precision} + \mathrm{recall}}$. 
The mean and the variance for this measure are estimated on the basis of the 20 tests and the 10-fold cross validation.

Figure \ref{fig:effetmixknn1} gives the mean and variance of the error rates for the 20 tests
%and F1 scores 
obtained when a 10 fold cross validation is performed using a 1-NN classifier. As shown in this figure the 1-NN classification using the \textit{induced} similarity measures alone performs the worse, which was expected since the variance on the comparability mapping is three times the one used to generate the native similarity measures. The error rate is thus 46\%in $\mathcal{S}$ and in $\mathcal{S}'$ when the \textit{induced} similarity measures alone are used and 35\% in  $\mathcal{S}$ and 39\% in $\mathcal{S}'$ when the \textit{native} similarity measures alone are used. 
%The F1 scores when native or induced similarity measures are used are close, around 39\% in $\mathcal{S}$ and 38\% in $\mathcal{S}'$. 

The effect of mixing \textit{native} and \textit{induced} similarity measures is strongly effective on these synthetic data sets since for both spaces the error rates drop below 25\% and reach a minimum when $\alpha=0.75$. Note that the variance of the mean error decrease slightly when the mixing parameter $\alpha$ is around this optimal value $0.75$. 
%The maximum F1 score is reached for $\alpha=0.8$.  
This experiment shows that even when the \textit{native} and \textit{induced} similarity measures are significantly noisy, the combination of the two sources of information allows for a significant reduction of the noise. 

This is precisely this effect that we would like to show on real bilingual comparable data, when the comparability mapping is elaborated from a bilingual lexicon.

\section{Variations around a quantitative comparability measure for bilingual texts}

\subsection{Comparability measure by Li and Gaussier ($C_{LG}$)}

The quantitative comparability measure proposed by \cite{LiGaussier2010} is based on the simple counting of \textit{word translation connections} that exist between two corpora in different languages according to a translation lexicon. Formally, let $\mathcal{S}_1$ and $\mathcal{S}_2$ be two corpora  expressed respectively in language $\mathcal{L}_1$  and $\mathcal{L}_2$. This comparability measure is formally defined as:

\begin{equation}
\label{Eq.1}
C_{LG}(\mathcal{S}_1,\mathcal{S}_2)=\displaystyle 
\frac{\displaystyle \sum_{w_1 \in W\mathcal{S}_1 \cap WD_1} \!\!\!\!\!\!\!\!\!\!\sigma(w_1)+
      \displaystyle \!\!\!\!\!\!\!\!\!\!\sum_{w_2 \in W\mathcal{S}_2 \cap WD_2} \!\!\!\!\!\!\!\!\!\!\sigma(w_2)}
      {|W\mathcal{S}_1 \cap WD_1|+|W\mathcal{S}_2 \cap WD_2|}
\end{equation}

where: $W\mathcal{S}_i$, $i \in \{1,2\}$ is the lexicon in language $\mathcal{L}_i$ associated with the corpus $\mathcal{S}_i$; $WD_i$ is the set of entries for language $\mathcal{L}_i$ into the bilingual dictionary that occur in $W\mathcal{S}_i$; $\sigma(w_i)$ is an indicator function that takes the value $1$ if at least one potential translation of the term  $w_i \in W\mathcal{S}_i$ in language $\mathcal{L}_i$ exists in the vocabulary associated with the corpus of the other language, $0$ otherwise.

This measure was originally designed for a bilingual lexicon extraction purposes, and not for the clustering or categorization of textual data. Hence, the authors did not incorporate any term weighting since it is \textit{a priori} irrelevant for a lexicon extraction task. However, if their definition is in line with a general definition of comparability such as the one given in introduction, the lack of term weighting is questionable when addressing a clustering/categorization task. The two variants that we propose hereinafter introduce a term weighting based on the number of term occurrences to specifically adapt the measure defined by  \cite{LiGaussier2010} to clustering or categorizing tasks.

\subsection{Enrichment of the $C_{LG}$ measure}

The $C_{LG}$ measure proposed by Li and Gaussier (eq.\ref{Eq.1}) takes account of neither the number of occurrences of the lexical entries in the documents nor their number of translations into the paired documents. The binary presence or absence of joint translation entries that is modeled by the indicator function $\sigma(w_i)$ is a strong feature that may affect the average comparability between pairs of documents. This could be the case when addressing corpora for which frequency of lexical entries helps discriminating between genres and topics. We propose the following two similar variants of the $C_{LG}$ measure that explicitly propose to go beyond the presence or absence of joint translations, conjecturing that this improvement will produce a positive effect in certain situations and tasks.

\subsubsection{First variant : $C_{VA_1}$}
 
The first variant symmetrically exploits (from the stand point of $\mathcal{L}_1$ and $\mathcal{L}_2$ languages) the following three elements: the number of occurrences of entries $w$ taken into the vocabulary of the first language corpus, the number of their translations in the bilingual dictionary and the presence of at least one of their translations in the vocabulary of the second language corpus.

Let $A_{1|2}$, $A_1$, $A_{2|1}$, $A_2$ be defined as follows:
 
\begin{eqnarray}
\begin{array}{ll}
A_{1|2}=\displaystyle \!\!\!\!\!\!\!\sum_{w_1 \in W\mathcal{S}_1 \cap WD_1}
\left(\frac{tf(w_1,\mathcal{S}_1)}{\tau(w_1,WD_1)} \cdot \sigma(w_1)\right) \nonumber\\
A_1=\displaystyle \!\!\!\!\!\!\!\sum_{w_1 \in W\mathcal{S}_1 \cap WD_1}
\left(\frac{tf(w_1,\mathcal{S}_1)}{\tau(w_1,WD_1)} \right)\\
A_{2|1}=\displaystyle \!\!\!\!\!\!\!\sum_{w_2 \in W\mathcal{S}_2 \cap WD_2}
\left(\frac{tf(w_2,\mathcal{S}_2)}{\tau(w_2,WD_2)} \cdot \sigma(w_2)\right)\\
A_2=\displaystyle \!\!\!\!\!\!\!\sum_{w_2 \in W\mathcal{S}_2 \cap WD_2}
\left(\frac{tf(w_2,\mathcal{S}_2)}{\tau(w_2,WD_2)} \right)\\
\end{array}
\end{eqnarray}

where $tf(w_i,\mathcal{S}_i)$ is the number of occurrences of entry $w_i$ in the corpus $\mathcal{S}_i$ expressed in language $\mathcal{L}_i$, $i \in \{1,2\}$; $\tau(w_i,WD_i)$ is the number of translations of entry $w_i$ of the corpus $\mathcal{S}_i$ in the dictionary $WD_i$;  $\sigma(w_i)$ is defined as above.

\begin{equation}
\label{Eq.2}
C_{VA_1}=\frac{1}{2} \cdot \left(\frac{A_{1|2}}{A_1}+\frac{A_{2|1}}{A_2}\right)\\
\end{equation}

\subsubsection{Second variant : $C_{VA_2}$}
This second variant is very similar to the previous one. It distinguishes mainly on the way the measure is symmetrized. Basically the first variant relates to a geometric mean while the second variant relates to an arithmetic mean.

\begin{equation}
\label{Eq.3}
C_{VA_2}= \frac{A_{1|2}+A_{2|1}}{A_1+A_2} \\
\end{equation}

\section{Experimenting on textual bilingual data}

We have collected the assessment corpora from 21 Wikipedia categories, from English (EN) and French (FR) languages. It originally consists of 154828 documents in total with 87793 English documents and 67035 French documents categorized in 21 categories, taken from existing Wikipedia categories. Since such corpus is thematically very large, corresponding similarity and comparability matrices are basically very sparse. To avoid  the algorithmic complexity behind the calculation of the induces similarity matrices $(O(N^{3}))$, we proceeded as follows which drastically reduces the sparsity of our matrices: \\
\begin{enumerate}
\item For each class and each language, we evaluate firstly the intra-language similarity matrices, using a cosine similarity based on a $tf-idf$ weighting,
\item secondly, we prune these intra-language similarity matrices using a threshold (typically 0.5) and order the documents according to their number of remaining neighbors (with whom they share a similarity above the threshold). 
\item by keeping for each language the best hundred documents, we get a refined comparable bilingual corpus.
\item Finally, to complexify the experiment, we enrich this corpus by adding, for each language, and for each class, 50\% of the initial number of documents. These added documents are randomly drawn from the initial 21 Wikipedia categories. 
\end{enumerate}
Each Wikipedia article is then represented by its plain textual content. tags and hyperlink have thus been removed.  
This Wikipedia corpus\footnote{The Wikipedia corpus is  available at \url{http://people.irisa.fr/Pierre-Francois.Marteau/Corpora/Wikipedia_21classes.zip}} contains 5822 documents in total, and is composed with 2745 French documents and 3077 English documents distributed into the 21 categories as listed in Table \ref{tab:cat}.
%For each class and each language, we have refined about one hundred documents by a strategy that we calculate similarities between the English documents to construct a similarities matrix, the we count for each line, the number of values which are greater than a high threshold (We have put 0.5) and find the highest ligne, after that, we stock in this line, the first biggest hundred documents. Then we enrich the refined corpus by adding ramdonly 50\% of the number of the documents for each class of each language. Therefore, finally, we have constructed this test corpus. 

\begin{table}[!h]
\center
\setlength{\tabcolsep}{-0pt}
\begin{tabular}{|l|c||l|c||l|c||l|c|}
  \hline
  EN categories &\# doc & FR categories &\# doc & EN categories &\# doc & FR categories &\# doc\\
  \hline
  \hline
 Astronomy &151 & Astronomie &123 & Movie &151 & Film &151\\
 Biology &151 & Biologie &115 & Music &151& Musique &151\\
 Economy &144& Economie&151 & Skating &151& Patinage &151\\
 Food &147 &Nourriture &4 & Heritage &151& Patrimoine &151\\
 Football &151 & Football &151 & Politics &151 & Politique &151\\
 Genetics &82& Génétique &151 & Religion &150 & Religion &133\\
 Geograpphy &139& Geographie &151 & Rugby &151 &Rugby &151\\
 Computer &151& Ordinateur &151&  Health &151 & Santé &63\\
 Literature &150 & Littérature &151 & Sculpture &151& Sculpture &151\\
 Mathematics &151 & Mathématique &63 & Tennis &151& Tennis &151\\
 Medicine &151 &Médecine &130 & & & &\\
 \hline
\end{tabular}
\caption{Composition of the comparable bilingual corpus extracted from Wikipedia (EN: English, FR: French)}
\label{tab:cat}
\end{table}

This corpus has been lemmatized using the TreeTagger \cite{schmid1994} \cite{TreeTagger2009}. Stoplists for French and English languages have been used and the term frequencies (\textit{tf}) for each vocabulary entry/document pair have been evaluated, as well as the inverse document frequencies \textit{idf} \cite{tfidf} that were estimated on the corpus.  Each Wikipedia article is finally represented by a \textit{tf-idf} weighted vector according to the classical vector space model \cite{Salton1975}.

\subsection{Bilingual dictionary}
To estimate the quantitative comparability between a pair of Enlish/French documents we have used the bilingual dictionary available at ELRA under reference ELRA-M0033. This dictionary contains 243,580 pairs of lexical entries in French and in English, which decompose into 110,541 lexical entries in English and 109,196 lexical entries in French.

The influence of the dictionary coverage rate has been partially studied in \cite{LiGaussier2010} and \cite{ke20014b}. It is shown that, for all three comparability measures $C_{LG}$, $C_{VA_1}$ and $C_{VA_2}$, the correlation of these measures with a gold standard comparability measure reference degrades when the dictionary coverage rate relatively to the corpus lexicon decreases. We do not address this issue in this paper, keeping in mind that an enrichment of the bilingual dictionary by including in particular domain dependent bilingual terminology entries would likely greatly improve our results.

\subsection{Evaluation measures}
The performance of the 1-NN classifier is evaluated using the classification error rate estimate using a 10-fold cross validation.
The performance of the tested clustering algorithms are also evaluated by comparing the predicted label for each document with its \textit{true} label. The accuracy (AC) and normalized mutual information (NMI) measures are used to evaluate the clustering performance \cite{XuLiuGong03}. As an internal evaluation scheme for estimating the quality of the clustering obtained in each linguistic space, we also use the Davies–Bouldin index (DB) \cite{DaviesBouldin1979} which roughly measures the quotient of intra and inter cluster average similarity measures.\\

The accuracy (AC) measure is defined as follows: it measures the fraction of documents that are correctly labeled, assuming a one-to-one correspondence between true categories and assigned clusters. Let $p$ denote any possible permutation of index set of clusters and \textit{true} categories. The Accuracy is thus defined as
\begin{eqnarray}
AC=\frac{1}{N}MAX_{p} \sum_{i=1 \cdots K} n_{i,p(i)} 
\end{eqnarray}
where $n_{i,p(i)}$ denotes the number of documents shared by class $i$ and cluster $p(i)$, $K$ is the number of categories and clusters, and $N$ is the total number of documents.\\

The $NMI$ measure between the \textit{true} clustering $\mathcal{C}$ and the predicted one $\mathcal{\tilde{C}}$ is defined as follows:
\begin{equation}
NMI(\mathcal{\tilde{C}}, \mathcal{C})=\frac{I(\mathcal{\tilde{C}}, \mathcal{C})}{(H(\mathcal{\tilde{C}})+H(\mathcal{C}))/2}
\end{equation}
with 
\begin{eqnarray}
I( \mathcal{\tilde{C}}, \mathcal{C} )
&=&
\sum_k \sum_j     P(\tilde{c}_k \cap c_j) \log\frac{P(\tilde{c}_k \cap c_j)}{P(\tilde{c}_k)P(c_j)}\nonumber
%&=&
%\sum_k \sum_j     \frac{|\tilde{c}_k \cap c_j|}{N} \log
%\frac{N|\tilde{c}_k \cap c_j|}{|\tilde{c}_k||c_j|}
\end{eqnarray}
and
\begin{eqnarray}
H(\mathcal{\tilde{C}}) &=& -\sum_k P(\tilde{c}_k) \log P(\tilde{c}_k)\nonumber\\
H(\mathcal{C}) &=& -\sum_k P(c_k) \log P(c_k)\nonumber
%&=& -\sum_k \frac{|\tilde{c}_k|}{N} \log \frac{|\tilde{c}_k|}{N}
\end{eqnarray}

The Davies-Boulding index DB is a data intrinsic evaluation measure, which is defined as follows
\begin{eqnarray}
DB = \frac {1} {K} \sum_{i=1}^{n} \max_{i\neq j}\left(\frac{\sigma_i + \sigma_j} {d(c_i,c_j)}\right) 
\end{eqnarray}
where $K$ is the number of clusters, $C_k$ is the centroid of cluster, $\sigma_k$ is the average distance of all elements in cluster $k$ to centroid $c_k$, and $d(c_i,c_j)$ is the distance between centroids $i$ and $j$.  The lower is this DB index value, the better is the clustering since this corresponds to low intra-cluster distances (high intra-cluster similarity) and high inter-cluster distances (low inter-cluster similarity).

\section{Experiments}
On the basis of the categorized comparable corpora collected from Wikipedia, we assess the benefit of mixing native similarity measures with comparability  on a 1-NN classification task and on a k-medoid clustering \cite{KaufmanRousseeuw87} \cite{KaufmanRousseeuw90} task. 

\subsection{1-NN classification task}

\begin{figure}[!h]
  \centering
  \begin{tabular}{ccc}
	\includegraphics[scale=0.45, angle=-90]{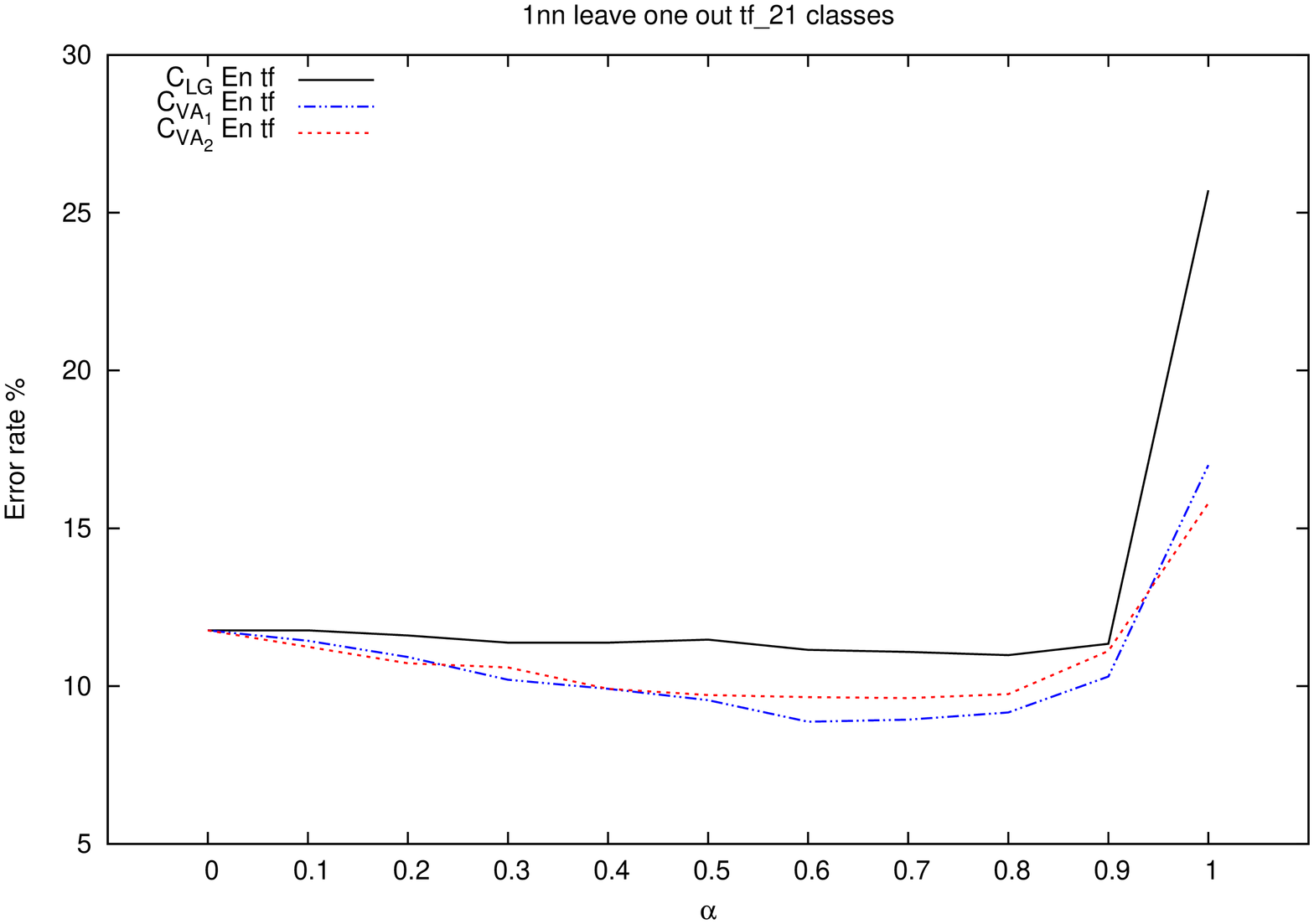}&\\
	\includegraphics[scale=0.45, angle=-90]{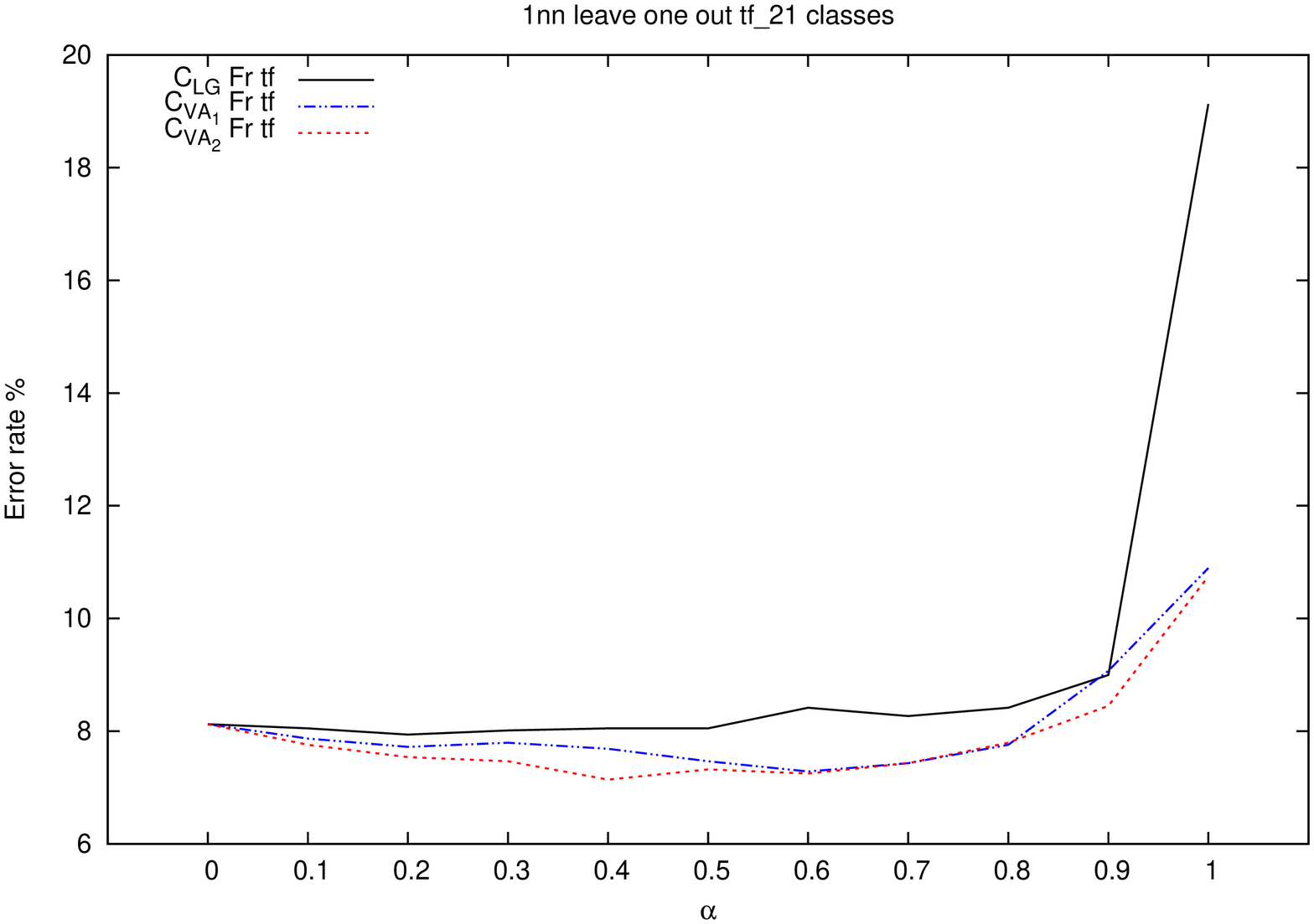}&
  
  \end{tabular}
  \caption{Comparability/similarity mixing effect on the 1-NN classification task, according to the leave one out error rate (top EN documents, bottom FR documents). $C_{LG}$ (black plain curve), $C_{VA_1}$ (blue dashdotted curve), $C_{VA_1}$ (red dotted curve) measures are given as a function of the mixing parameter $\alpha$.}
	\label{fig:effetcompknn1}
\end{figure}

\begin{figure}[!h]
  \centering
  \begin{tabular}{ccc}
	\includegraphics[scale=0.45, angle=-90]{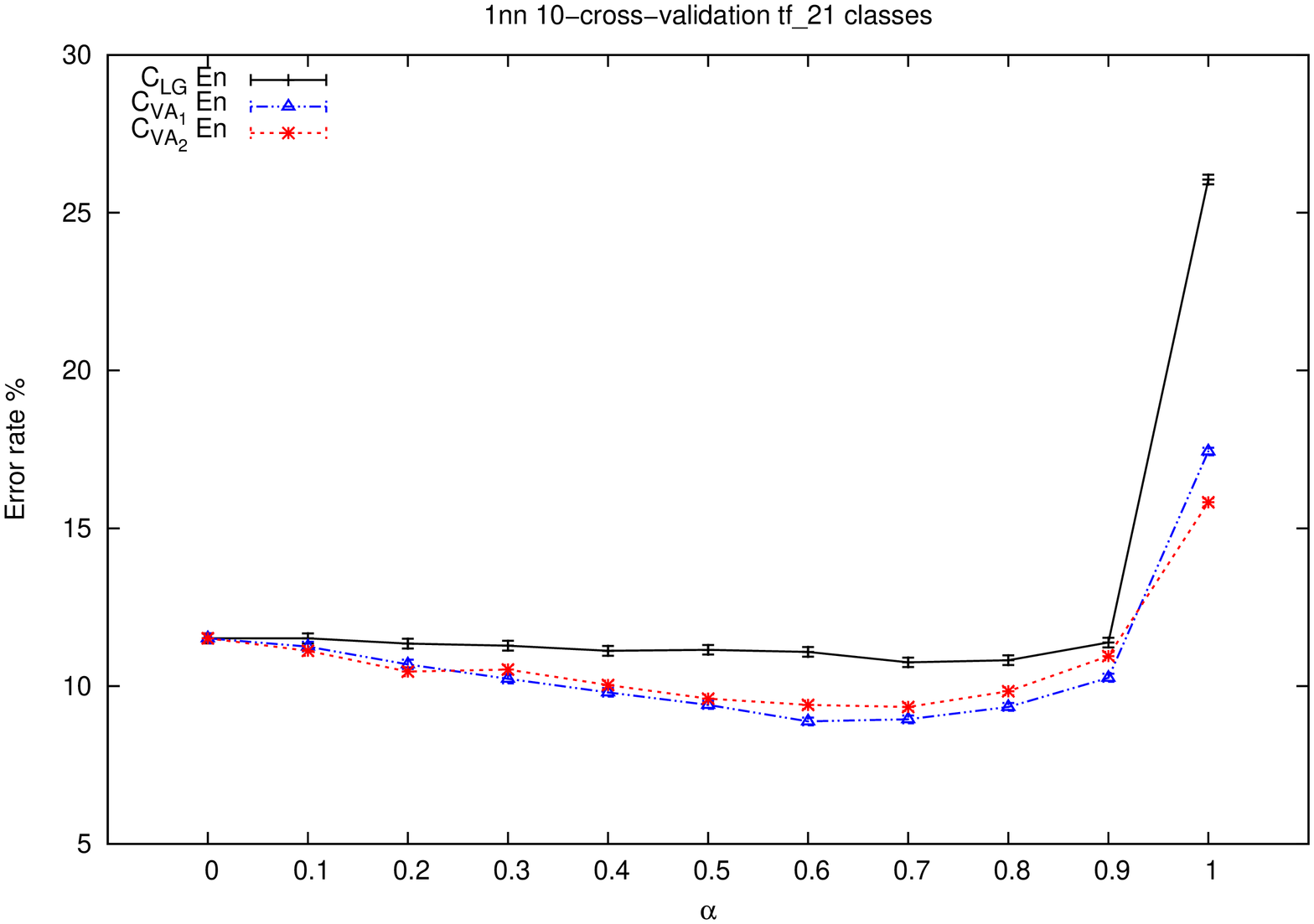}&\\
	\includegraphics[scale=0.45, angle=-90]{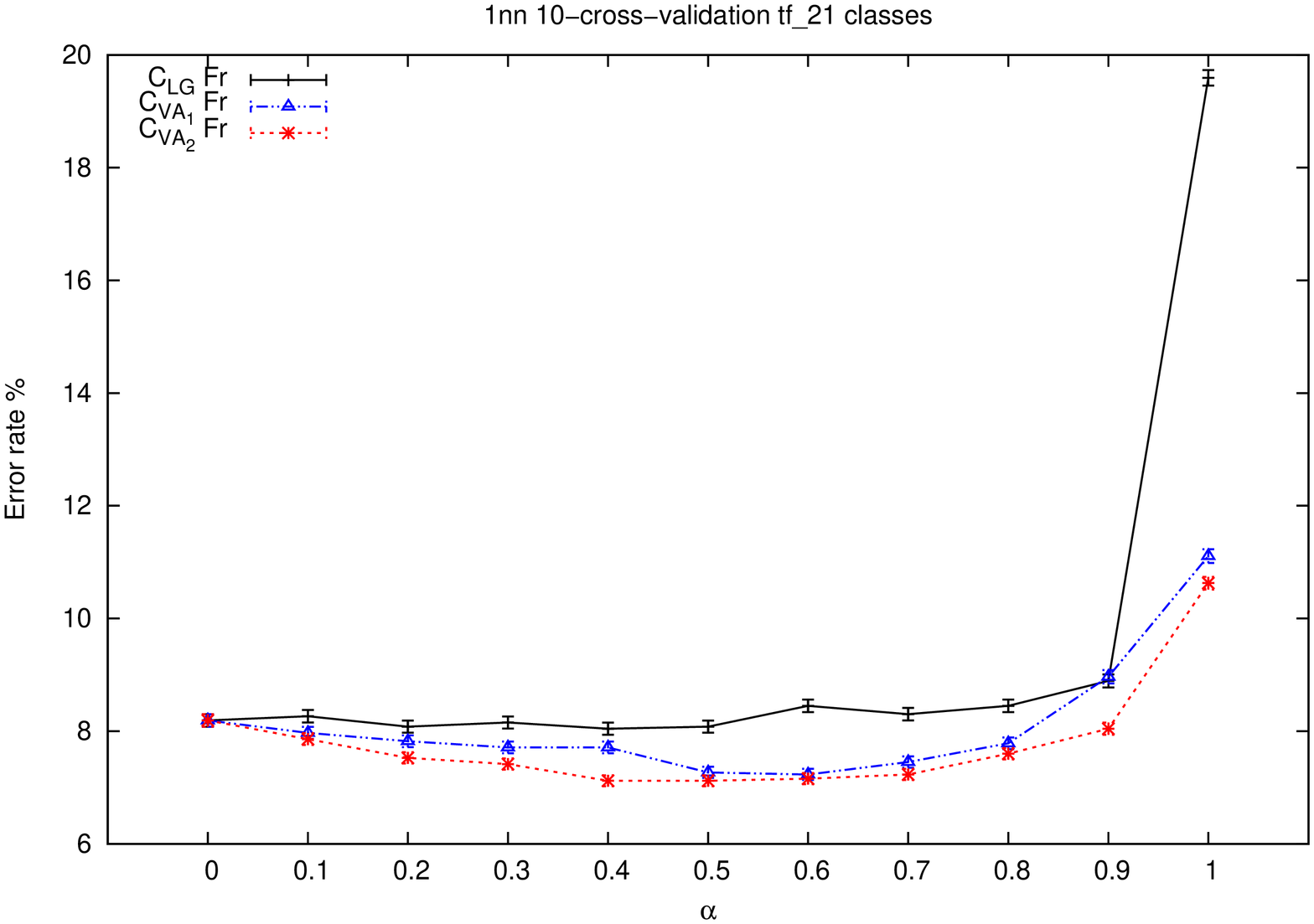}&
  
  \end{tabular}
  \caption{Comparability/similarity mixing effect on the 1-NN classification task, according to $10$-fold cross validation error rate (top EN documents, bottom FR documents). $C_{LG}$ (black plain curve), $C_{VA_1}$ (blue triangle dashdotted curve), $C_{VA_1}$ (red star dotted curve) measures are given as a function of the mixing parameter $\alpha$.}
	\label{fig:effetcompknn2}
\end{figure}
We first study the effect of mixing similarity and comparability on the 1-NN classification task while varying the parameter $\alpha \in [0,1]$.\\

Figures \ref{fig:effetcompknn1} and  \ref{fig:effetcompknn2} show that the similarity/comparability mixing has a significant impact for the two variants $C_{VA_1}$ and $C_{VA_2}$ since it allows reducing by 3\% the error rate of the classification for the English language documents and 1.5\% for the French language documents. However, comparatively, the $C_{LG}$ measure improves slightly for both languages the classification accuracy, and is less stable when $\alpha$ varies.

%\FloatBarrier

\subsection{k-medoids clustering task}

\begin{figure}[]
  \centering
  \begin{tabular}{cc}
    \includegraphics[angle=270, scale=0.45]{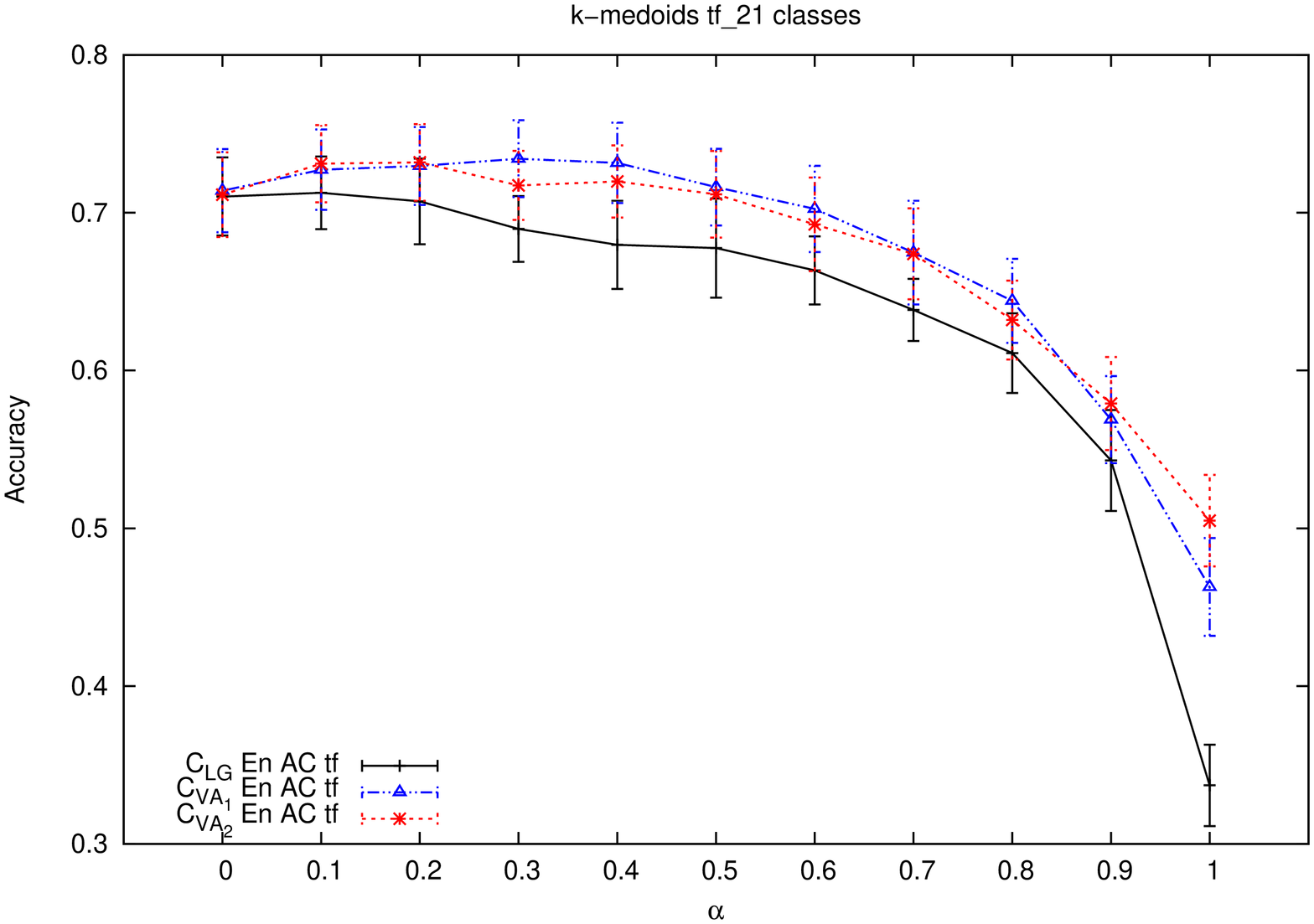} \\
	\includegraphics[angle=270, scale=0.45]{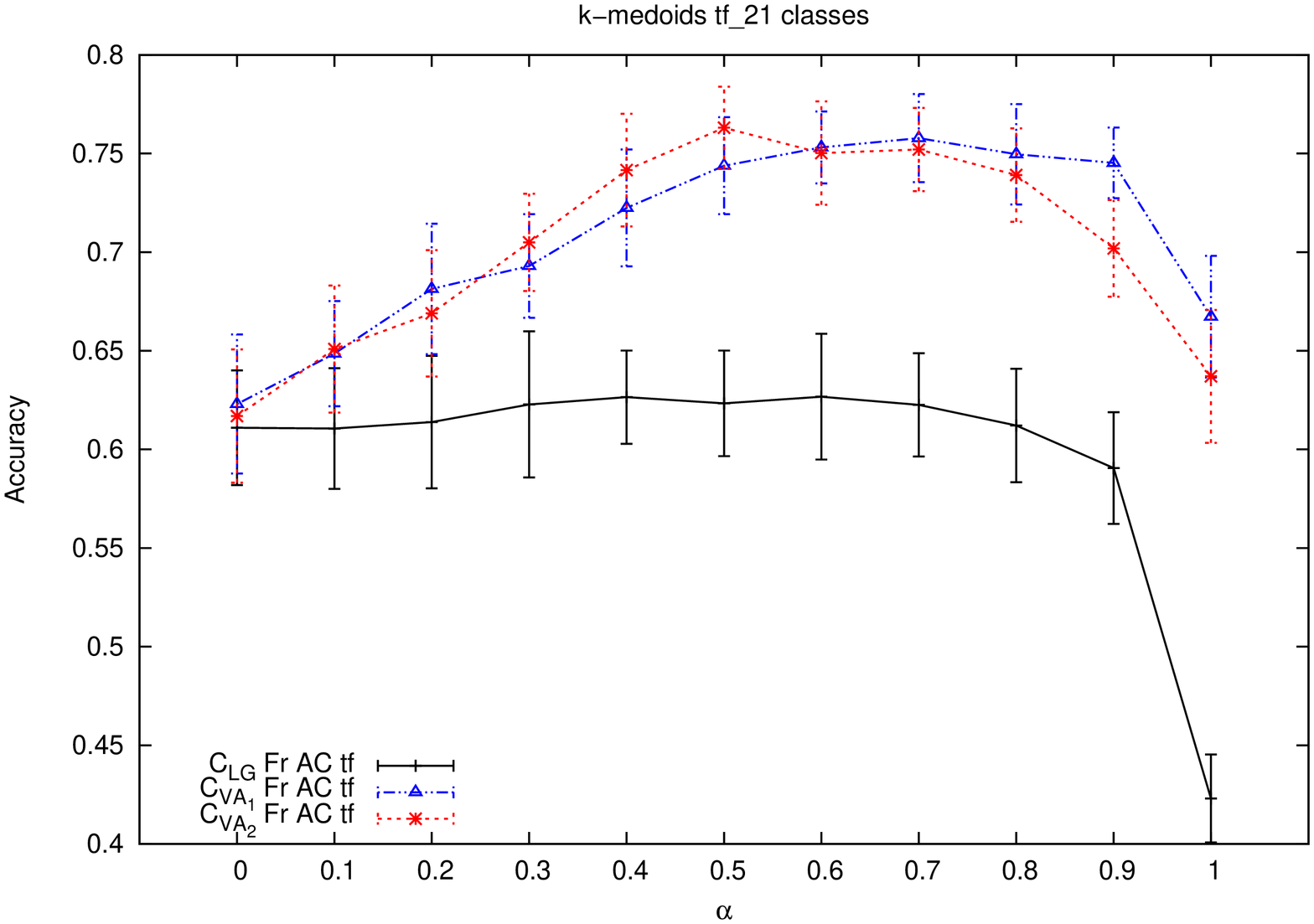} \\
  \end{tabular}
  \caption{Evaluation of the comparability/similarity mixing on the k-medoids clustering accuracy (AC) (top EN documents, bottom FR documents). $C_{LG}$ (black plain curve), $C_{VA_1}$ (blue triangle dashdotted curve), $C_{VA_1}$ (red star dotted curve) measures are given as a function of the mixing parameter $\alpha$.}
\label{fig:effetac}
\end{figure}

\begin{figure}[!h]
 \centering
  \begin{tabular}{cc}
    \includegraphics[angle=270, scale=0.45]{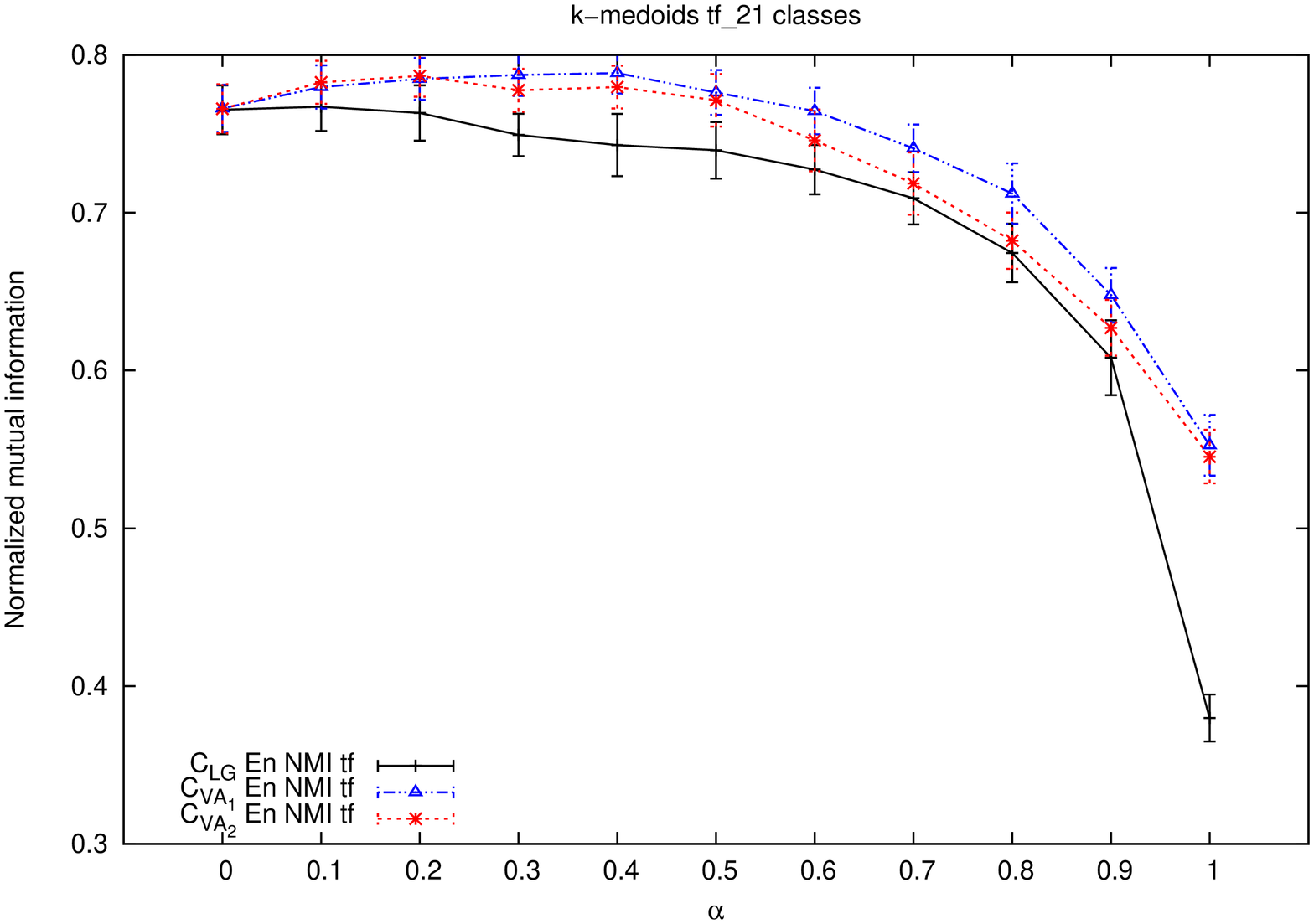} \\
	\includegraphics[angle=270, scale=0.45]{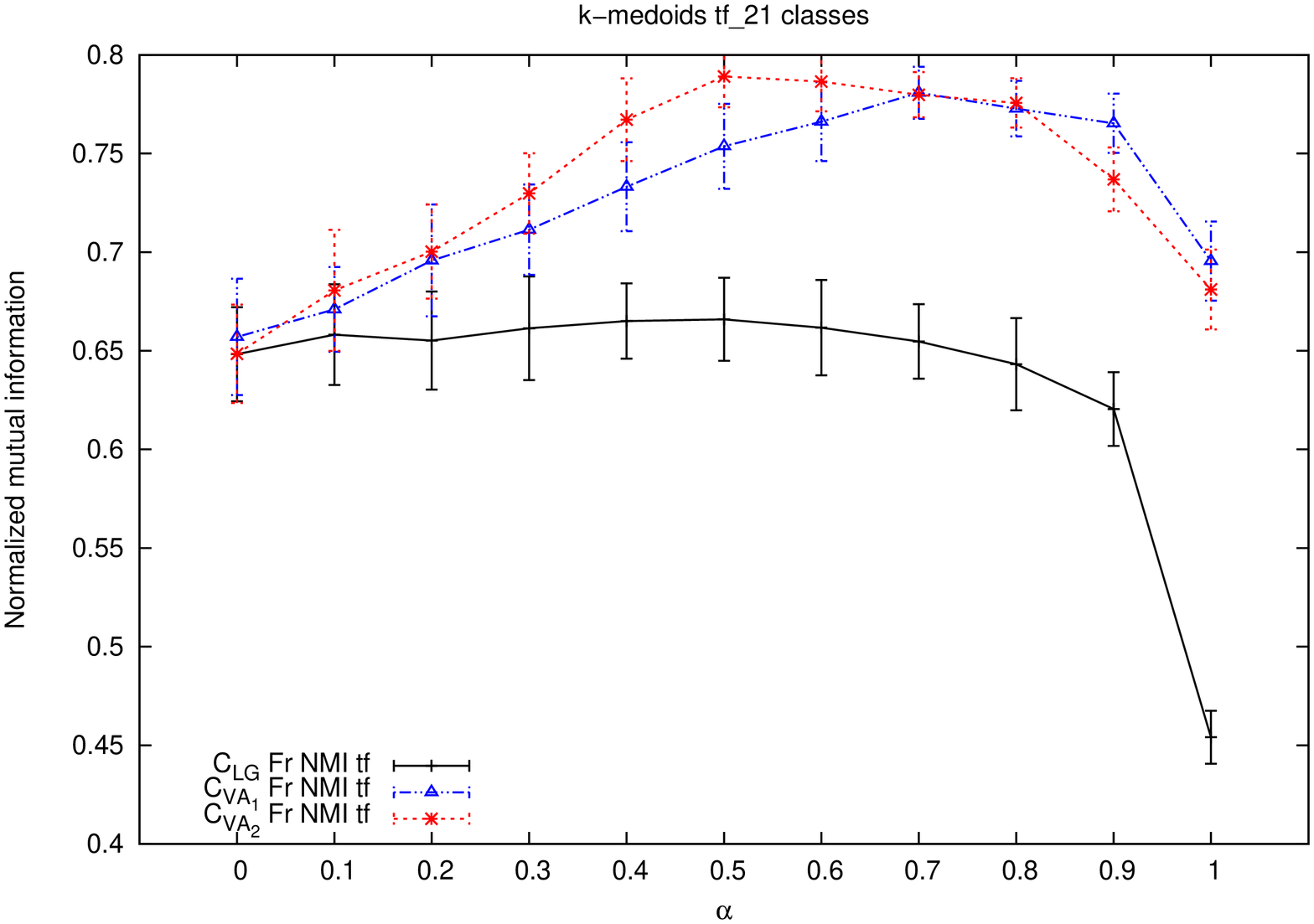} \\
  \end{tabular}
  \caption{Evaluation of the mixing of comparability and similarity measures on the k-medoids clustering according to the NMI measure (top EN documents, bottom FR documents). $C_{LG}$ (black plain curve), $C_{VA_1}$ (blue triangle dashdotted curve), $C_{VA_1}$ (red star dotted curve) measures are given as a function of the mixing parameter $\alpha$.}
\label{fig:effetnmi}
\end{figure}

\begin{figure}[!h]
  \centering
  \begin{tabular}{cc}
    \includegraphics[angle=270, scale=0.45]{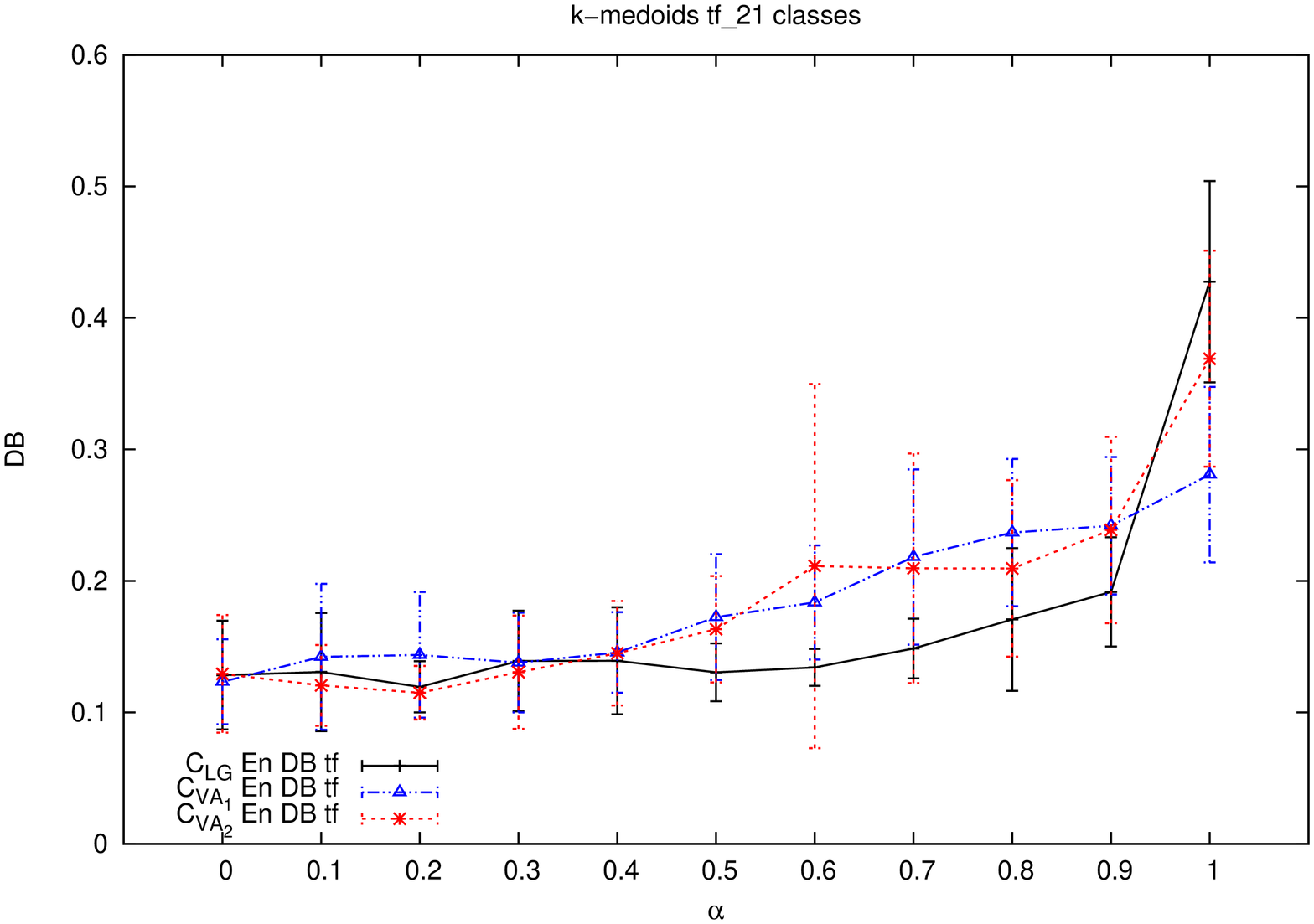} \\
	\includegraphics[angle=270, scale=0.45]{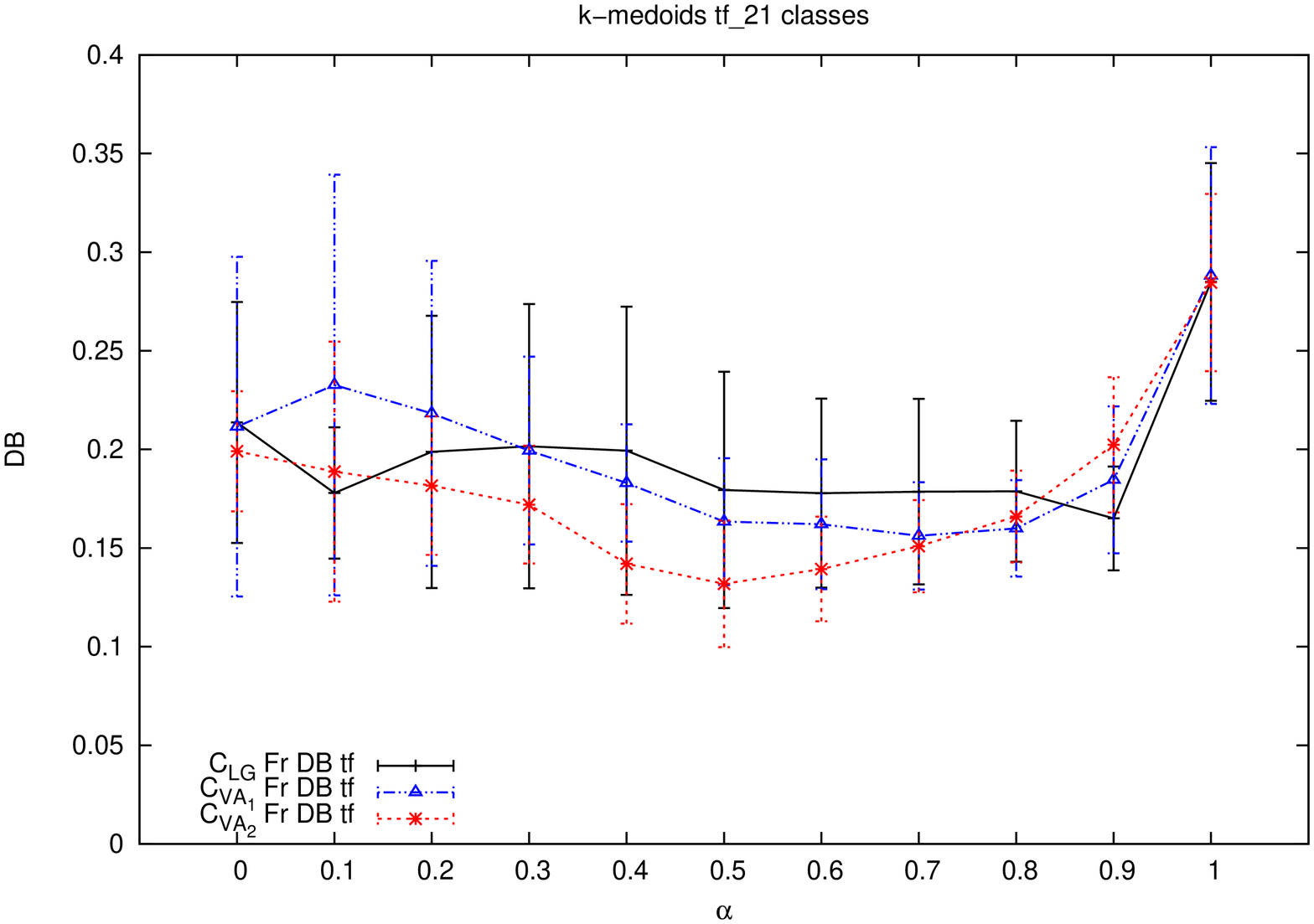} \\
  \end{tabular}
  \caption{Comparability/similarity mixing effect on a k-medoids clustering according to the DB measure (top EN documents, bottom FR documents). $C_{LG}$ (black plain curve), $C_{VA_1}$ (blue triangle dashdotted curve), $C_{VA_1}$ (red star dotted curve) measures are given as a function of the mixing parameter $\alpha$.}
\label{fig:effetintrainter}
\end{figure}

We study here the effect of mixing comparability and similarity measures on a k-medoids clustering task for all three comparability measures. We used the previously defined AC, NMI and DB measures for the assessment of this clustering task.\\

 Figures \ref{fig:effetac} and \ref{fig:effetnmi} show that both AC and NMI measures can be improved up to 15\% in the scope of the clustering of French language documents and up to 3\% in the scope of the clustering of English language documents for both $C_{VA_1}$ and $C_{VA_2}$ measures. However, once again, the $C_{LG}$ brings comparatively less improvement for both languages. \\

Figure \ref{fig:effetintrainter} depicts the DB measure as a function of parameter $\alpha$, for all three comparability measures. It is shown that, for $C_{VA_1}$ and $C_{VA_2}$, this ratio decreases for some \textit{good} $\alpha$ values, especially for the French language, whereas for the measure $C_{LG}$, this value increases in general. A \textit{good} mixing of the comparability and similarity measures has thus a positive impact when using $C_{VA_1}$ and $C_{VA_2}$ measures and a rather negative impact when using the $C_{LG}$ measure. \\

\section{Analysis and conclusions}

In this paper, we have proposed a new approach for the co-clustering and co-categorization of bi-lingual data when a comparability mapping exists. This approach, that could be characterized as a kind of three-mode clustering or categorization, is based on the concept of similarity \textit{induced} by a comparability bipartite graph. The three-mode data analysis scheme is implemented as a mixing model used to merge \textit{native} and \textit{induced} similarity measures inside each of the two linguistic space. The assessment of this mixing model on purely synthetic random data is quite informative and demonstrates the noise reduction capability of the method. 

On real bilingual textual data, the approach involves a quantitative comparability measure that is based on the exploitation of a bilingual dictionary. To this end, two variants of the comparability measure proposed by \cite{LiGaussier2010} have been proposed to adapt this measure to clustering and categorization tasks. The implementation of our model on semi-manually constructed comparable corpora collected from the Wikipedia resource shows to be quite effective. Our detailed experimentation shows that the mixing of \textit{native} similarity measures with a quantitative comparability measure has a clear impact on the classification and clustering accuracies. It is noticeable that the improvement is more important in the French linguistic space comparatively to the English linguistic space. Furthermore, our approach works specifically well for the $C_{VA_1}$ and $C_{VA_2}$ comparability variant measures with stable and robust classification or clustering result improvements.

It nevertheless has a small positive impact when the $C_{LG}$ measure is used, leading to conclude that taking into account of the frequency of  occurrence of lexical entries and frequencies of their translations into the comparability measure design is of crucial importance for thematic classification or clustering of bilingual English/French documents. One potential explanation is that these frequencies of occurrence pair well with the \textit{tf-idf} heuristic that takes place in native \textit{cosine} similarity. Moreover, according to our results, the choice of the value of the mixing parameter $\alpha$ is quite important. A relatively high  $\alpha$ value (between 0.5 and 0.8), that slightly favors the \textit{induced} similarity measures comparatively to the \textit{native} similarity,  will be a good choice in general. Finally our experimentation shows that the $C_{VA_2}$, whose symmetrization is homogeneous to an arithmetic mean, is more robust than $C_{VA_1}$, a result that needs to be consolidated on other independent experiments.

In terms of perspective, ensuring the scalability and generalizing the approach and experimentation are major prospects to help constructing thematic comparable corpora on demand. 

The bilingual dictionary is a particularly important resource in our approach, since the quality of the comparability mapping linking the two linguistic spaces directly relies on it. The impact of the coverage of the dictionary relatively to the corpus has been partly studied in \cite{LiGaussier2010} and \cite{ke20014b}. In the context of comparable thematic data processing, it is likely that the enrichment of a general bilingual resource by introducing domain specific terminology entries would bring some benefit. 

Finally, another perspective is to expand it to various pairing of languages for which bilingual resources are available, in particular bilingual dictionaries.

%%================================================================
\section*{Acknowledgements} 
This work has been partially funded by the French National Research Agency (ANR-METRICC).

%%================================================================
%% Note : si l'on préfère éviter de factoriser les crossrefs :
%% bibtex -min-crossrefs 99 taln-exemple
%%================================================================

%\bibliography{biblio}
%\bibliographystyle{plain}

\bibliography{biblio}

\begin{thebibliography}{}

\bibitem[\protect\BCAY{Amini\ \BBA\ Goutte}{Amini\ \BBA\
  Goutte}{2010}]{AminiGoutte2010}
Amini, M.-R.\BBACOMMA\  \BBA\ Goutte, C. \BBOP2010\BBCP.
\newblock \BBOQ A co-classification approach to learning from multilingual
  corpora\BBCQ\
\newblock {\Bem Machine Learning}, {\Bem 79\/}(1-2), 105--121.

\bibitem[\protect\BCAY{Baker}{Baker}{1996}]{Baker1996}
Baker, M. \BBOP1996\BBCP.
\newblock \BBOQ Corpus-based translation studies: The challenges that lie
  ahead.\BBCQ\
\newblock In {\Bem In Terminology, LSP and Translation: Studies in Language
  Engineering in Honour of Juan C. Sager}.

\bibitem[\protect\BCAY{Davies\ \BBA\ Bouldin}{Davies\ \BBA\
  Bouldin}{1979}]{DaviesBouldin1979}
Davies, D.~L.\BBACOMMA\  \BBA\ Bouldin, D.~W. \BBOP1979\BBCP.
\newblock \BBOQ A cluster separation measure\BBCQ\
\newblock {\Bem Pattern Analysis and Machine Intelligence, IEEE Transactions},
  {\Bem PAMI-1(2)}, 224--227.

\bibitem[\protect\BCAY{Déjean\ \BBA\ Gaussier}{Déjean\ \BBA\
  Gaussier}{2002}]{Dejean2002}
Déjean, H.\BBACOMMA\  \BBA\ Gaussier, E. \BBOP2002\BBCP.
\newblock \BBOQ Une nouvelle approche a l’extraction de lexiques bilingues à
  partir de corpus comparables\BBCQ\
\newblock {\Bem Lexicometrica}, {\Bem Numéro spécial, corpus alignés},
  1--22.

\bibitem[\protect\BCAY{EAGLES}{EAGLES}{1996}]{EAGLES1996}
EAGLES \BBOP1996\BBCP.
\newblock \BBOQ Expert advisory group on language engineering standards
  guidelines: http://www.ilc.pi.cnr.it/eagles96/browse.html\BBCQ\
\newblock \BTR, EAGLES.

\bibitem[\protect\BCAY{Fung\ \BBA\ Yee}{Fung\ \BBA\ Yee}{1998}]{Fung1998}
Fung, P.\BBACOMMA\  \BBA\ Yee, L.~Y. \BBOP1998\BBCP.
\newblock \BBOQ An ir approach for translating new words from nonparallel,
  comparable texts\BBCQ\
\newblock In {\Bem Proc. of the 36th ACL meeting, Vol. 1}, ACL '98, \BPGS\
  414--420, Stroudsburg, PA, USA. ACL.

\bibitem[\protect\BCAY{Jagarlamudi, Daum{\'e},\ \BBA\ Udupa}{Jagarlamudi
  et~al.}{2011a}]{Jagarlamudi:2011ACL}
Jagarlamudi, J., Daum{\'e}, III, H., \BBA\ Udupa, R. \BBOP2011a\BBCP.
\newblock \BBOQ From bilingual dictionaries to interlingual document
  representations\BBCQ\
\newblock In {\Bem Proc. ACL-HLT - Vol. 2}, HLT '11, \BPGS\ 147--152,
  Stroudsburg, PA, USA. ACL.

\bibitem[\protect\BCAY{Jagarlamudi, Udupa, Daum{\'e},\ \BBA\ Bhole}{Jagarlamudi
  et~al.}{2011b}]{Jagarlamudi:2011EMNLP}
Jagarlamudi, J., Udupa, R., Daum{\'e}, III, H., \BBA\ Bhole, A.
  \BBOP2011b\BBCP.
\newblock \BBOQ Improving bilingual projections via sparse covariance
  matrices\BBCQ\
\newblock In {\Bem Proc.s of the Conf. on EMNLP}, \BPGS\ 930--940, Stroudsburg,
  PA, USA. Association for Computational Linguistics.

\bibitem[\protect\BCAY{Kaufman\ \BBA\ Rousseeuw}{Kaufman\ \BBA\
  Rousseeuw}{1987}]{KaufmanRousseeuw87}
Kaufman, L.\BBACOMMA\  \BBA\ Rousseeuw, P.~J. \BBOP1987\BBCP.
\newblock {\Bem Clustering by means of Medoids, in Statistical Data Analysis
  Based on the $L_1$–Norm and Related Methods}.
\newblock North-Holland.

\bibitem[\protect\BCAY{Kaufman\ \BBA\ Rousseeuw}{Kaufman\ \BBA\
  Rousseeuw}{1990}]{KaufmanRousseeuw90}
Kaufman, L.\BBACOMMA\  \BBA\ Rousseeuw, P.~J. \BBOP1990\BBCP.
\newblock {\Bem Finding groups in data: an introduction to cluster analysis}.
\newblock John Wiley and Sons, New York.

\bibitem[\protect\BCAY{Ke, Marteau,\ \BBA\ M{\'e}nier}{Ke
  et~al.}{2014}]{ke20014b}
Ke, G., Marteau, P.-F., \BBA\ M{\'e}nier, G. \BBOP2014\BBCP.
\newblock \BBOQ {Variations on quantitative comparability measures and their
  evaluations on synthetic French-English comparable corpora}\BBCQ\
\newblock In {\Bem {LREC 2014, the 9th edition of the Language Resources and
  Evaluation Conference}}, \BPG~pp, Reykjavik, Iceland.

\bibitem[\protect\BCAY{Li\ \BBA\ Gaussier}{Li\ \BBA\
  Gaussier}{2010}]{LiGaussier2010}
Li, B.\BBACOMMA\  \BBA\ Gaussier, E. \BBOP2010\BBCP.
\newblock \BBOQ Improving corpus comparability for bilingual lexicon extraction
  from comparable corpora\BBCQ\
\newblock In {\Bem COLING}, \BPGS\ 644--652.

\bibitem[\protect\BCAY{Li, Gaussier,\ \BBA\ Aizawa}{Li
  et~al.}{2011}]{LiGaussier2011}
Li, B., Gaussier, E., \BBA\ Aizawa, A. \BBOP2011\BBCP.
\newblock \BBOQ Clustering comparable corpora for bilingual lexicon
  extraction\BBCQ\
\newblock In {\Bem Proc. of the 49th ACL-HLT- Vol. 2}, \BPGS\ 473--478,
  Stroudsburg, PA, USA. Association for Computational Linguistics.

\bibitem[\protect\BCAY{Marteau\ \BBA\ Ménier}{Marteau\ \BBA\
  Ménier}{2013}]{MarteauMenier2013}
Marteau, P.-F.\BBACOMMA\  \BBA\ Ménier, G. \BBOP2013\BBCP.
\newblock \BBOQ Similarités induites par mesure de comparabilité :
  signification et utilité pour le clustering et l’alignement de textes
  comparables.\BBCQ\
\newblock In {\Bem TALN}, \BPG\ 515–522.

\bibitem[\protect\BCAY{Mirkin}{Mirkin}{1996}]{MirkinBoris96}
Mirkin, B. \BBOP1996\BBCP.
\newblock {\Bem Mathematical Classification and Clustering}.
\newblock Kluwer Academic Publishers.

\bibitem[\protect\BCAY{Munteanu, Fraser,\ \BBA\ Marcu}{Munteanu
  et~al.}{2004}]{MunteanuFM04}
Munteanu, D.~S., Fraser, A., \BBA\ Marcu, D. \BBOP2004\BBCP.
\newblock \BBOQ Improved machine translation performance via parallel sentence
  extraction from comparable corpora\BBCQ\
\newblock In {\Bem HLT-NAACL}, \BPGS\ 265--272.

\bibitem[\protect\BCAY{Salton, Wong,\ \BBA\ Yang}{Salton
  et~al.}{1975}]{Salton1975}
Salton, G., Wong, A., \BBA\ Yang, C.~S. \BBOP1975\BBCP.
\newblock \BBOQ A vector space model for automatic indexing\BBCQ\
\newblock {\Bem Commun. ACM}, {\Bem 18\/}(11), 613--620.

\bibitem[\protect\BCAY{Schmid}{Schmid}{1994}]{schmid1994}
Schmid, H. \BBOP1994\BBCP.
\newblock \BBOQ {Probabilistic Part-of-Speech Tagging Using Decision
  Trees}\BBCQ\
\newblock In {\Bem Proceedings of the Int. Conf.e on New Methods in Language
  Processing}, \BPGS\ 44--49.

\bibitem[\protect\BCAY{Schmid}{Schmid}{2009}]{TreeTagger2009}
Schmid, H. \BBOP2009\BBCP.
\newblock \BBOQ {TreeTagger,
  www.ims.uni-stuttgart.de/projekte/corplex/TreeTagger/}\BBCQ.

\bibitem[\protect\BCAY{{Sp\"{a}rck Jones}}{{Sp\"{a}rck Jones}}{1972}]{tfidf}
{Sp\"{a}rck Jones}, K. \BBOP1972\BBCP.
\newblock \BBOQ A statistical interpretation of term specificity and its
  application in retrieval\BBCQ\
\newblock {\Bem Journal of Documentation}, {\Bem 28}, 11--21.

\bibitem[\protect\BCAY{Van Mechelen~I}{Van
  Mechelen~I}{2004}]{VanMechelenBockDeBoeck04}
Van Mechelen~I, Bock~HH, D. B.~P. \BBOP2004\BBCP.
\newblock \BBOQ Two-mode clustering methods:a structured overview\BBCQ\
\newblock {\Bem Statistical Methods in Medical Research}, {\Bem 13(5)},
  363--394.

\bibitem[\protect\BCAY{Vu, Aw,\ \BBA\ Zhang}{Vu et~al.}{2009}]{Vu:2009}
Vu, T., Aw, A.~T., \BBA\ Zhang, M. \BBOP2009\BBCP.
\newblock \BBOQ Feature-based method for document alignment in comparable news
  corpora\BBCQ\
\newblock In {\Bem Proceedings of the 12th EACL Conf.}, \BPGS\ 843--851,
  Stroudsburg, PA, USA. ACL.

\bibitem[\protect\BCAY{Wei~Xu\ \BBA\ Gong}{Wei~Xu\ \BBA\
  Gong}{2003}]{XuLiuGong03}
Wei~Xu, X.~L.\BBACOMMA\  \BBA\ Gong, Y. \BBOP2003\BBCP.
\newblock \BBOQ Document clustering based on non-negative matrix
  factorization\BBCQ\
\newblock In {\Bem SIGIR'03}, \BPGS\ 267--273.

\end{thebibliography}
\bibliographystyle{theapa}

%%================================================================
\end{document}